\documentstyle[aps,pra,amsfonts,preprint]{revtex}
 \tighten
\begin{document}
\draft
\title{Primality Test Via Quantum Factorization}
\author{H. F. Chau\footnote{Present address: Department of Physics, University 
 of Hong Kong, Pokfulam Road, Hong Kong. E-mail: hfchau@hkusua.hku.hk} ~and 
 H.-K. Lo\footnote{Present address: BRIMS, Hewlett-Packard Labs, Filton Road,
 Stoke Gifford, Bristol, Bs12 6QZ, U. K. E-mail: hkl@hplb.hpl.hp.com}}
\address{
 School of Natural Sciences, Institute for Advanced Study, Olden Lane,
 Princeton, NJ 08540
}
\date{\today}
\preprint{IASSNS-HEP-95/69; quant-ph:9508005}
\maketitle
\begin{abstract}
 We consider a probabilistic quantum implementation of a variation of the
 Pocklington-Lehmer $N-1$ primality test using Shor's algorithm. $\mbox{O}(
 \log^3 N \log\log N \log\log\log N )$ elementary q-bit operations are required
 to determine the primality of a number $N$, making it (asymptotically) the
 fastest known primality test. Thus, the potential power of quantum mechanical
 computers is once again revealed.
\end{abstract}
\pacs{\noindent
 \begin{minipage}[t]{5in}
   \vspace{1ex}
  \\
  \begin{tabular}{ll}
   AMS 1991 subject classification: & (Primary) 11Y05, 11Y11, 68Q10, 81V99 \\
   & (Secondary) 11A51, 11Y16, 68Q20
  \end{tabular}
  \vspace{0.9ex} \\
  \begin{tabular}{ll}
   Keywords: & Computational Complexity, Pocklington-Lehmer $N-1$ Primality
   Test,\\
   & Quantum Computation, Quantum Factorization, Shor's Algorithm
  \end{tabular}
 \end{minipage}
}
\vspace{12ex}
]
\par
 Finding large primes and factorizing large composite numbers are two classic
 mathematical problems of great practical interest. For instance, in the RSA
 public key cryptography, the key, which is made public, is the product of two
 large primes whose values are kept secret. The secret values of the two primes
 are needed to decode the encoded messages (ciphertexts). The security of this
 scheme lies in the difficulty in factoring large composites. More concretely,
 while multiplying two integers can be done in a time polynomial in the number
 of digits of the two integers (and hence ``efficient''), the fastest
 factorization algorithm that runs on classical computers (or Turing machines)
 takes almost an exponential amount of time ($\sim \exp(L^{1/3})$ where $L$ is
 the number of digits of the number to be factorized) \cite{Public_Key}.
 Consequently, given the value of the public key, it is almost hopeless for an
 eavesdropper to attempt to break the RSA cryptographic scheme by factoring the
 key into two large primes. For this reason, finding an efficient factorization
 method is the dream of eavesdroppers. On the other hand, for additional
 security of the RSA scheme, the public key has to be changed frequently to
 avoid ``accidental'' factorization of the key. To fulfill this need, an
 efficient algorithm for proving the primality of a large integer is required.
\par
 The possibility of performing classical computation by using quantum
 mechanical machines has been investigated by various people
 \cite{Ben,Feynman_1,Feynman_2,Deutsch_1,Deutsch_2}. Recently, Shor discovered
 an efficient quantum factorization algorithm \cite{Shor,RMP}: By using the
 massive parallelism and interference effect in quantum mechanics, which have
 no classical counterparts, Shor found an efficient method to compute the
 period of a function. This method immediately leads to efficient algorithms
 for both the discrete logarithm and factorization problems \cite{Shor}.
 Therefore, if a quantum mechanical computer is ever built, the RSA
 crypto-system will no longer be secure. Some people have even proposed that
 quantum cryptography will ultimately be the only way to ensure the security
 of a cryptosystem \cite{Bennett:92,Jozsa:94,Schumacher:95,Proc}.
\par
 Primality tests\footnote{A {\it primality} test is an algorithm which outputs
 ``true'' if and only if the input is a prime. It may not halt if the input is
 composite. This is to be distinguished from a {\it compositeness} test that
 may occasionally indicate a number as prime even when it is in fact
 composite.} are generally much easier than factorization. The APRCL test
 (based on Jacobi sum) is one of the most commonly used algorithms. The number
 of elementary bit operations needed for testing the primality of a number $N$
 is $\mbox{O}((\log N)^{c\log\log\log N})$ for some constant $c > 0$
 \cite{APR}. Although the run time of this algorithm is not truly polynomial in
 $\log N$, it works reasonably fast for numbers of less than 1,000 decimal
 digits \cite{Cohen}. The first polynomial time probabilistic primality test
 was proposed by Goldwasser and Kilian \cite{Goldwasser} using ideas from
 elliptic curves. Their algorithm was later implemented by Atkin and Morain
 \cite{Cohen,Atkin}. Its run time scales as $\mbox{O}(\log^6 N)$. However,
 their algorithm assumes some unproven (although very plausible) conjectures in
 analytic number theory \cite{Atkin} and may fail to work for an infinite
 sequence of (non-random) prime numbers \cite{AH:92} even though it will never
 mis-identify a composite number as a prime. Finally, using ideas from Abelian
 varieties, Adleman and Huang \cite{AH:92} discovered a polynomial time
 probabilistic primality test without any unproven hypothesis. However, their
 algorithm is extremely complicated and is totally impractical to implement
 \cite{Cohen}.
\par
 Further improvements may be possible. In fact, if we assume that validity of
 the (yet still unproven) Extended Riemann Hypothesis, then there is a
 deterministic primality test whose run time scales as $\mbox{O}((\log N)^4
 \log\log\log N)$ \cite{ERH}. More recently, there is statistical evidence
 supporting the conjecture that the primality of a number $N$ can be proven
 deterministically in $\mbox{O}((\log N)^3 \log\log N \log\log\log N)$ time
 \cite{ERH2}.
\par
 In this paper, we propose a straightforward probabilistic primality test
 based on the Pocklington-Lehmer $N-1$ method using the quantum factorization
 algorithm. Its run time scales as $\mbox{O}((\log N)^3 \log\log N \log\log\log
 N )$ and is thus asymptotically faster than all known classical primality
 tests. In the discussion below, we will always assume that the number $N$ has
 failed all common compositeness tests and hence is very likely to be a prime
 (see, for example, Refs.~\cite{Public_Key,Cohen,Riesel} for some simple and
 efficient compositeness tests).
\par
 Note that the number of primitive residue classes ($\mbox{mod~} N$) is $N-1$
 and the multiplicative group formed by the primitive residue classes has a
 generator of order $N-1$ if and only if $N$ is a prime \cite{Riesel}. This
 leads us to the following theorem:
\par \medskip \noindent
{\it Theorem~1:} (Pocklington-Lehmer $N-1$ test) Suppose $N-1 = \prod_{j=1}^{m}
 p_j^{\beta_j}$ with all $p_j$'s distinct primes. If there exists $a\in
 {\Bbb Z}_N$ such that
\begin{equation}
 \left\{ \begin{array}{l}
 a^{(N-1)/p_j} \not\equiv 1 ~~(\mbox{mod~} N) \hspace{0.2in} \mbox{for~} j = 
 1,2,\ldots ,m \vspace{1em} \\
 a^{N-1} \equiv 1 ~~(\mbox{mod~} N)
 \end{array} \right. \mbox{~,}
\label{E:Lehmer_Test}
\end{equation}
then $N$ is a prime \cite{Cohen,Riesel}.
\par \medskip
 Thus, the test consists of two parts, namely, the complete factorization of
 the number $N-1$, and the verification of conditions in
 Eq.~(\ref{E:Lehmer_Test}). Since the test requires the complete factorization
 of a number, it is not a good general purpose primality test before the
 discovery of Shor's quantum factorization algorithm. (As mentioned earlier, no
 efficient classical factorization method is known. The fastest known classical
 method for factorization is the number field sieve. Under reasonable heuristic
 assumptions, it takes $\mbox{O}(\exp (c (\log M)^{1/3} (\log\log M)^{2/3}))$
 elementary operations for some constant $c > 0$ \cite{NFS} to find a factor of
 a number $M$.)
\par
 The situation is completely different after Shor's discovery. As shown in the
 Appendix~A, Shor's algorithm requires $\mbox{O}((\log M)^2 (\log\log M)^2
 \log\log\log M)$ elementary q-bit operations\footnote{A quantum mechanical bit
 is now commonly called a ``q-bit''. Loosely speaking, coherent superposition
 of states allows a q-bit to hold more information than a classical bit. In
 addition, ``elementary'' here refers to operation in the form of unitary
 operator acting on one or two q-bits. Please refer to
 Refs.~\cite{Bennett,Sleator,HF} for constructions of ``elementary'' logical
 operators.} to find a factor of a composite number $M$, provided that $M$ is
 not in the form of $p^n$ or $2p^n$ for some odd prime number $p$. In the case
 that $M$ is of the form $p^n$ for an odd prime $p$, there exists a classical
 algorithm to find $p$ and $n$ in $\mbox{O}((\log M)^2 (\log\log M)^2
 \log\log\log M)$ time \cite{Cohen}. Alternatively, we show in Appendix~B that
 this can be done equally efficiently by using a quantum algorithm similar to
 Shor's algorithm. Since factorization of a prime power is much easier than
 that of a composite number with distinct prime factors, we shall only consider
 the latter in our computational complexity analysis.
\par
 Let us consider the first part of the test --- the complete factorization of
 the number $N-1$. Suppose we would like to factorize $M \equiv N-1$
 completely. Using Shor's algorithm, we can find a non-trivial factor $f$ of
 $M$ in $\mbox{O}((\log M)^2 (\log\log M)^2 \log\log\log M)$ elementary
 operations. The problem then reduces to the factorization of the numbers $f$
 and $M/f$. We can further speed up the process by extracting multiple factors
 of $M$, if any, by computing $\gcd (f, M/f)$. Clearly, this takes negligible
 time as compared to the Shor's algorithm. And the complete factorization of
 $M$ is obtain by recursively applying Shor's algorithm $m-1$ times where $m$
 is the number of distinct primes of $M$. Since the product of the first $m$
 primes is of order of $2^m$ \cite{Riesel}, so complete factorization of $M$
 requires the running of Shor's algorithm for at most $\mbox{O}(\log M)$ times.
 Thus, no more than $\mbox{O}((\log M )^3 (\log\log M)^2 \log\log\log M)$
 elementary operations are needed for running Shor's algorithm alone. In
 addition, we also need to verify that a {\em complete factorization} of $M
 \equiv N-1$ has been obtained. That is, the $p_j$'s we have found in
 Theorem~1 are indeed prime numbers. Let us denote the number of elementary
 operations needed for the first and second parts of the primality test for a
 $M$ by $P_1(M)$ and $P_2(M)$ respectively. Also, let $P(M) = P_1(M) + P_2(M)$.
 From the above discussion,
\begin{eqnarray}
 & & P_1(N) \nonumber \\
 & \leq & \sum^m_{i=1} P(p_i) + \mbox{O} ((\log N)^3 (\log\log N)^2
 \log\log\log N) \mbox{~.}
 \label{first}
\end{eqnarray}
\par\indent
 Let us come to the second part of the test --- the application of the
 Pocklington-Lehmer $N-1$ test. We choose an integer $m$ randomly and test if
 all the conditions in Eq.~(\ref{E:Lehmer_Test}) are satisfied. Now there are
 at most $\mbox{O} (\log N)$ such conditions. Using the power algorithm
 \cite{Knuth}, verification of each condition in Eq.~(\ref{E:Lehmer_Test})
 requires $\mbox{O} (\log N)$ multiplications. Using the Sch\"{o}nhangen and
 Strassen method, multiplying two number of size at most $N$ can be done in
 $\mbox{O} (\log N \log\log N \log\log\log N)$ elementary operations
 \cite{Knuth,Fast_Mult}. Therefore, altogether $\mbox{O} ((\log N)^3 \log\log N
 \log\log\log N)$ elementary operations are needed for each random number $m$
 chosen. It can be shown that the probability that a randomly chosen integer
 $m$ satisfying all the conditions in Eq.~(\ref{E:Lehmer_Test}) is at least
 $\mbox{O} ( 1/ \log\log N)$ \cite{Hardy}. Consequently, the total number of
 elementary operations needed for the second part of the test is given by
\begin{equation}
 P_2(N) \leq \mbox{O} ((\log N)^3 (\log\log N)^2 \log\log\log N) \mbox{~.}
 \label{second}
\end{equation}
\par\indent 
 Notice that if $\Pi_{i=1}^{m} p_i^{\beta_i}$ is the prime number decomposition
 of a positive integer $N$ (with $p_i$ are distinct primes), then
\begin{mathletters}
\begin{equation}
 \sum_{i=1}^{m} \log p_i \leq \log N \mbox{~,}
\end{equation}
\begin{equation}
 \sum_{i=1}^{m} \log\log p_i \leq \log\log N \mbox{~,}
\end{equation}
 and hence
\begin{eqnarray}
 \sum_{i=1}^{m} (\log p_i)^3 \leq \left( \sum_{i=1}^{m} \log p_i \right)^3 \leq
 (\log N)^3 \mbox{~.}
\end{eqnarray}
\end{mathletters}
 Therefore,
\begin{eqnarray}
 & & \sum_{i=1}^m (\log p_i)^3 (\log\log p_i)^2 (\log\log\log p_i) \nonumber \\
 & \leq & (\log N)^3 (\log\log N)^2 (\log\log\log N) \mbox{~.}
 \label{product}
\end{eqnarray}
 By induction, it is straightforward to prove from
 Eqs.~(\ref{first})--(\ref{product}) that
\begin{eqnarray}
 P(N) & = & P_1(N) + P_2(N) \nonumber \\
 & \leq & \mbox{O} ((\log N)^3 (\log\log N)^2 \log\log\log N) \mbox{~.}
 \label{all}
\end{eqnarray}
\par\indent
 Note that if $N$ is in fact a composite number, this primality test will never
 terminate. The Pocklington-Lehmer test gives a certificate for primality once
 the number $N$ passes it. On the other hand, Shor's algorithm is efficient for
 finding non-trivial factors of a number. Thus, Shor's algorithm and our
 quantum primality test are complimentary to each other.
\par
 Although Eq.~(\ref{all}) already tells us that the above quantum primality
 test algorithm is already better than all the classical algorithms known to
 date, we now go on to describe a fine tuning of our quantum algorithm which
 reduces the run time by a factor of $\log\log N$. As shown in the operation
 counting analysis above, both the quantum factorization and the verification
 of Eq.~(\ref{E:Lehmer_Test}) are equally fast. Thus, in order to reduce the
 run time of the quantum Pocklington-Lehmer algorithm, we have to speed up both
 parts.
\par
 To speed up the quantum factorization, we can perform trial divisions to
 eliminate all the prime factors of $N-1$ that are smaller than $k$. This can
 be done by $\approx k$ divisions, taking $\mbox{O}(k \log N \log\log N
 \log\log\log N)$ time \cite{Fast_Mult}. After the trial division, we can
 concentrate on the prime factors of $N-1$ that are greater than $k$. Clearly,
 at most $\mbox{O}(\log N / \log k)$ distinct prime factors of $N-1$ are
 greater than $k$. So by combining the trial division with Shor's algorithm,
 $N-1$ can be factorized completely in $\mbox{O}(\log N \log\log N \log\log\log
 N (k + (\log N)^2 \log\log N / \log k))$ time. Optimal solution is obtained
 when we take the number of trial divisions $k \sim (\log N)^2 / \log\log N$.
 Therefore, factorization of $N-1$ requires only $\mbox{O}((\log N)^3 \log\log
 N \log\log\log N)$ elementary q-bit operations. (In case we have a prime
 number table up to the number $k$, then prime number theorem tells us that
 only $\mbox{O}(k / \log k)$ trial divisions are required. Using the same
 argument, we know that optimal solution occurs when $k \sim \log^2 N$.
 However, the optimal number of elementary q-bit operations is still $\mbox{O}
 ((\log N)^3 \log\log N \log\log\log N)$. That is, only a constant factor speed
 up is gained when we use a prime number table.)
\par
 To speed up the verification of primality, we employ a variation of the
 Pocklington-Lehmer test by Brillhart {\em et al.} \cite{Riesel,Brillhart}:
\par\medskip\noindent
{\it Theorem~2:} (Brillhart {\em et al.}) Suppose $N-1 = \prod_{j=1}^{m}
 p_j^{\beta_j}$ with all $p_j$'s distinct primes. And if, for each $j = 1,2,
 \ldots ,m$, there exists $a_j \in {\Bbb Z}_N $ such that
\begin{equation}
 \left\{ \begin{array}{l}
 a_j^{(N-1)/p_j} \not\equiv 1 ~~(\mbox{mod~} N) \vspace{1em} \\
 a_j^{N-1} \equiv 1 ~~(\mbox{mod~} N)
 \end{array} \right. \mbox{~,}
\label{E:Alternative}
\end{equation}
then $N$ is a prime.
\par\medskip
 Once again, we randomly choose an integer $m$ and test if it satisfies
 Eq.~(\ref{E:Alternative}). For each $p_j$, the probability that a randomly
 chosen $m$ satisfies the constraint $m^{(N-1)/p_j} \not\equiv 1 ~(\mbox{mod~}
 N)$ in Eq.~(\ref{E:Alternative}) is at least 1/2. Thus, for each $m$, half of
 the constraints in Eq.~(\ref{E:Alternative}) are satisfied on average. Thus,
 we are almost sure to have found all the required $a_j$ by randomly picking
 $m$'s and checking Eq.~(\ref{E:Alternative}) a few times. Obviously, our new
 verification process takes $\mbox{O}((\log N)^3 \log\log N \log\log\log N)$
 time. Combining the quantum factorization with trial division, and Theorem~2,
 we have a $\mbox{O}((\log N)^3 \log\log N \log\log\log N)$ run time quantum
 primality test as promised.
\par
 In summary, we have presented a probabilistic quantum primality test using a
 variation of the Pocklington-Lehmer $N-1$ test and Shor's quantum
 factorization algorithm. Its run time scales as $\mbox{O}((\log N)^3 \log\log
 N \log\log\log N)$. (Moreover, it requires $\mbox{O}(\log^2 N)$ bits of extra
 working space.) As far as we know, this is the (asymptotically) fastest
 primality test to date. Our quantum primality test can be further speeded up
 by a constant factor if we replace Theorem~2 by another variation of the
 Pocklington-Lehmer algorithm which involve only a partial factorization of
 $N-1$ (see Ref.~\cite{Riesel}, for example).
\par
 It is interesting to know if there exist an even faster primality test. In
 particular, if the conjecture by Bach and Huelsbergen is correct, then there
 is a deterministic primality test, whose run time is as good as ours (i.e.
 $\mbox{O}((\log N)^3 \log\log N \log\log\log N)$ \cite{ERH2}.
\acknowledgements{This work is supported by the DOE grant DE-FG02-90ER40542.}
\appendix
\section{Shor's Algorithm}
 We outline the idea of Shor's algorithm below (See Refs.~\cite{Shor,RMP} for
 details). To factorize a composite number $M$ (which is assumed not to be a
 prime power), we prepare our system in the state
\begin{equation}
 |\Psi\rangle = \frac{1}{2^{L/2}} \sum_{a=1}^{2^L} |a\rangle \mbox{~,}
 \label{E:A1}
\end{equation}
 with $2^L \approx M^2$. This can be achieved by, say, setting $L$ quantum
 spin-1/2 particles with their spins pointing towards the positive
 $x$-direction (and all our measurements are performed in the $z$-direction).
\par
 Then we evolve our wavefunction to
\begin{equation}
 |\Psi\rangle = \frac{1}{2^{L/2}} \sum_{a=1}^{2^L} |a,m^a \mbox{mod~} M\rangle
 \mbox{~,} \label{E:A2}
\end{equation}
 for some randomly chosen integer $1<m<M$ with $\gcd (m,M) = 1$. (If $\gcd
 (m,M) > 1$, then we are so lucky that we have found a non-trivial factor of
 $M$ by chance. The probability for this to happen scales exponentially with
 $\log M$, and is therefore negligible.) The above evolution can be done using
 the power algorithm \cite{Knuth}, which takes $\mbox{O}(L)$ multiplications in
 ${\Bbb Z}_M$. As mentioned in the text, multiplying two $L$-bit numbers using
 the Sch\"{o}nhagen and Strassen method, which is asymptotically the fastest
 known algorithm, requires $\mbox{O}(L \log L \log\log L)$ elementary bit
 operations \cite{Knuth,Fast_Mult}. Consequently, evolving the wavefunction
 from state~(\ref{E:A1}) to state~(\ref{E:A2}) takes $\mbox{O}(L^2 \log L
 \log\log L)$ elementary q-bit operations. Besides, it requires $\mbox{O}(L)$
 extra q-bits as working space during the computation.
\par
 Now we make a measurement on the second set of q-bits in our system (which
 should take at most $\mbox{O}(L)$ time). Thus, the  wavefunction of our system
 for the first set of q-bits collapses to
\begin{equation}
 |\Psi\rangle = \frac{1}{\sqrt{k}} \sum_{a=0}^{k} |a_0 + p a\rangle \mbox{~,}
\end{equation} 
 where $p$ is the order of the number $m$ under multiplication modulo $M$,
 $0\leq a_0 < k$ is some constant, and $k = \left[ (2^L - a_0) / p \right]$.
\par
 To extract the order $p$, we perform a discrete Fourier transform, which
 evolves our system to
\begin{equation}
 |\Psi\rangle = \frac{1}{\sqrt{k s^L}} \sum_{c=0}^{2^L -1} \sum_{a=0}^{k} \exp
 \left[ \frac{2\pi i (a_0 + p a) c}{2^L} \right] |c\rangle \mbox{~.}
\end{equation}
 This can be done in $\mbox{O}(L^2)$ elementary q-bit operations \cite{RMP}.
 Now the amplitude of our wavefunction is sharply peaked at $|p\rangle$. It can
 be shown that by making a measurement on the first set of q-bits, we have a
 probability of at least $\mbox{O}(1 / \log L)$ of getting the correct order
 $p$ \cite{Shor,RMP}. So, by repeatedly running our machine $\mbox{O}( \log L)$
 times, we are almost sure to get the order $p$ of the multiplicative group
 modulo $M$ generated by the integer $m$. Therefore, it requires $\mbox{O}(
 (\log M)^2 (\log\log M)^2 \log\log\log M)$ elementary q-bit operations to find
 the order $p$ of the group $\langle m\rangle$.
\par
 Now we hope that $p$ is an even number and that $\gcd ((m^{p/2}-1) \mbox{mod~}
 M, M)$ is a non-trivial factor of $M$. It can be shown that for a randomly
 chosen $m$, the probability that the above algorithm does give a non-trivial
 factor of $M$ is at least 1/2 provided that $M$ is not of the form $p^k$ or
 $2p^k$ for some odd prime $p$ \cite{RMP}.
\par
 The remaining case is to recognize and factorize an odd number $M$ in the form
 of a prime power. This can be done by classical probabilistic algorithms
 whose run time scales like $\mbox{O}((\log M)^2 (\log\log M)^2 \log\log\log
 M)$ \cite{Cohen}, which is negligible in comparison with Shor's algorithm.
 (See Appendix~B for an equally efficient quantum prime power factorization
 algorithm.) Thus, we have an efficient way to factorize a composite number
 $M$.
\par
 Combining Shor's algorithm with the classical factorization of prime powers,
 we are almost sure to find a non-trivial factor of $M$ after $\mbox{O}((\log M
 )^2 (\log\log M)^2 \log\log\log M)$ elementary q-bit operations. Moreover, the
 power algorithm is one of the major bottle necks in this method.
\section{Quantum Prime Power Factorization Algorithm}
 Here we discuss a variant of Shor's algorithm that is useful for factoring a
 number $M$ that is of the form $p^n$ for some odd prime $p$. Following Shor,
 we can find the order of a number $m$ which is relatively prime to $M \equiv
 p^n$ in $\mbox{O}((\log M)^2 (\log\log M)^2 \log\log\log M)$ time. We denote
 the set of all integers in ${\Bbb Z}_M$ which are relatively prime to $M$ by
 $U({\Bbb Z}_M)$. It can be shown that $U({\Bbb Z}_M)$ is a cyclic group of
 order $p^{n-1} (p-1)$ under multiplication modulo $M$ \cite{MNT}. The group
 generated by $m$ under multiplication modulo $M$, $\langle m\rangle$, is a
 sub-group of $U({\Bbb Z}_M)$. The probability that the order of $\langle m
 \rangle$ is divisible by $p^{n-1}$ equals the probability that a randomly
 chosen element of ${\Bbb Z}_{p^{n-1} (p-1)}$ is relatively prime to $p^{n-1}$,
 which is in turn equal to $1 - 1/p \geq 2/3$. So, the greatest common divisor
 of the order of $m$ and $M$ has at least 2/3 chance of being $p^{n-1}$. Thus,
 we have a probability of at least 2/3 of finding $p$ by calculating $M / \gcd
 (M,r)$ where $r$ is the order of $m$. Once $p$ is found, $M$ can be factorized
 easily. The total time required scales as $\mbox{O}((\log M)^2 (\log\log M)^2
 \log\log\log M)$.

\end{document}